\documentclass[
    ,final            
  ]
  {aipproc}

\layoutstyle{6x9}

\def\gsim{\mathrel{\hbox{\rlap{\lower.55ex \hbox {$\sim$}}
                   \kern-.3em \raise.4ex \hbox{$>$}}}}
\def\lsim{\mathrel{\hbox{\rlap{\lower.55ex \hbox {$\sim$}}
                   \kern-.3em \raise.4ex \hbox{$<$}}}}

\begin{document}

\title{The Radio Properties of Type Ibc Supernovae}


\author{Alicia M. Soderberg}{
  address={Caltech Astronomy, MC 105-24, Pasadena, CA 91106, USA}
}



\begin{abstract}
Over the past few years, long-duration $\gamma$-ray bursts (GRBs),
including the subclass of X-ray flashes (XRFs), have been revealed to
be a rare variety of Type Ibc supernova (SN Ibc).  While all these
events result from the death of massive stars, the electromagnetic
luminosities of GRBs and XRFs exceed those of ordinary Type Ibc SNe by
many orders of magnitude.  The observed diversity of stellar death
corresponds to large variations in the energy, velocity, and geometry
of the explosion ejecta.  Using multi-wavelength (radio, optical,
X-ray) observations of the nearest GRBs, XRFs, and SNe Ibc, I show
that while GRBs and XRFs couple at least $\sim 10^{48}$ erg to relativistic
material, SNe Ibc typically couple $\lsim 10^{48}$ erg to their
fastest (albeit non-relativistic) outflows.  Specifically, I find that
less than 3\% of local SNe Ibc show any evidence for association with
a GRB or XRF.  
Recently, a new class of GRBs and XRFs has been revealed which 
are under-luminous in comparison with the statistical sample of GRBs.
Owing to their faint high-energy emission, these sub-energetic bursts
are only detectable nearby ($z\lsim 0.1$) and are likely 10 times more
common than cosmological GRBs.  In comparison with local SNe Ibc and
typical GRBs/XRFs, these explosions are intermediate in terms of both 
volumetric rate and energetics.  Yet the essential physical process
that causes a dying star to produce a GRB, XRF, or sub-energetic
burst, and not just a SN, remains a crucial open question.  Progress
requires a detailed understanding of ordinary SNe Ibc which will be
facilitated with the launch of wide-field optical surveys in the near 
future.
\end{abstract}

\maketitle


\section{Introduction}

The discovery of several gamma-ray bursts (GRBs) and X-ray flashes
(XRFs) at $z\lsim 0.3$ in the last few years has firmly established
that GRBs and XRFs are accompanied by supernovae of Type Ibc (SNe Ibc)
(see \citealt{wb06} for a review).  While the distribution of
GRB/XRF-SN optical luminosities appears indistinguishable from that of
local SNe Ibc \citep{skp+06}, spectroscopy reveals unusually broad
absorption lines (``broad-lined'', BL) in every case \citep{pmm+06}.
At other wavelengths, these explosions are easily distinguished since
GRBs and XRFs produce mildly-relativistic ejecta, which gives rise to
strong non-thermal ``afterglow'' emission.  Radio observations are
critical in this analysis since they provide the best calorimetry of
the fastest ejecta.

Recently, we have identified a population of GRBs/XRFs that are
sub-energetic by a factor of $\sim 10^2$ and about 10 times more
common than typical bursts \citep{skb+04b,skn+06}.  Given their
under-luminous prompt energy release, current satellites can only
detect them nearby ($z\lsim 0.1$).  These sub-energetic explosions are
intermediate between GRBs/XRFs and local SNe Ibc with
respect to gamma-ray energy, jet collimation, and volumetric rate, and
thus hint at an overall continuum.

Motivated by the GRB/XRF-SN connection and the discovery of
sub-energetic GRBs, I embarked on a survey of
optically-selected local ($d\lsim 200$ Mpc) SNe Ibc with the Very
Large Array (VLA).  The goal is to determine the fraction of SNe Ibc that
produce mildly-relativistic ejecta. This survey is well-suited to
recognizing bursts for which no gamma-ray emission is detected, either
because the jets are pointed away from our line-of-sight
(``off-axis'') or the emission is below the detection thresholds of
current satellites.

Since then I have observed $\sim 200$ SNe Ibc with the VLA, on
timescales of days to years after the explosion.  This dedicated
effort has served to characterize in a systematic way the environments
and ejecta properties of SNe Ibc for the first time.  Early radio
observations are crucial for identifying sub-energetic and/or
uncollimated explosions.  On the other hand, late-time radio
observations are sensitive to jets initially beamed away from our
line-of-sight.  Through this intense VLA program, I have established
that (i) roughly 15\% of SNe Ibc show detectable radio emission, (ii)
less than $\sim 3\%$ show radio luminosities comparable to those
observed for GRB/XRF afterglows \citep{bkf+03,skn+06}, (iii) less than
10\% of SNe Ibc harbor typical GRBs pointed away from our
line-of-sight \cite{sfw04,snb+06}, and (iv) basic optical properties
(peak luminosity, photospheric velocities) are not reliable indicators
of strong radio emission and/or relativistic ejecta.

\begin{figure}
 \includegraphics[height=.5\textheight]{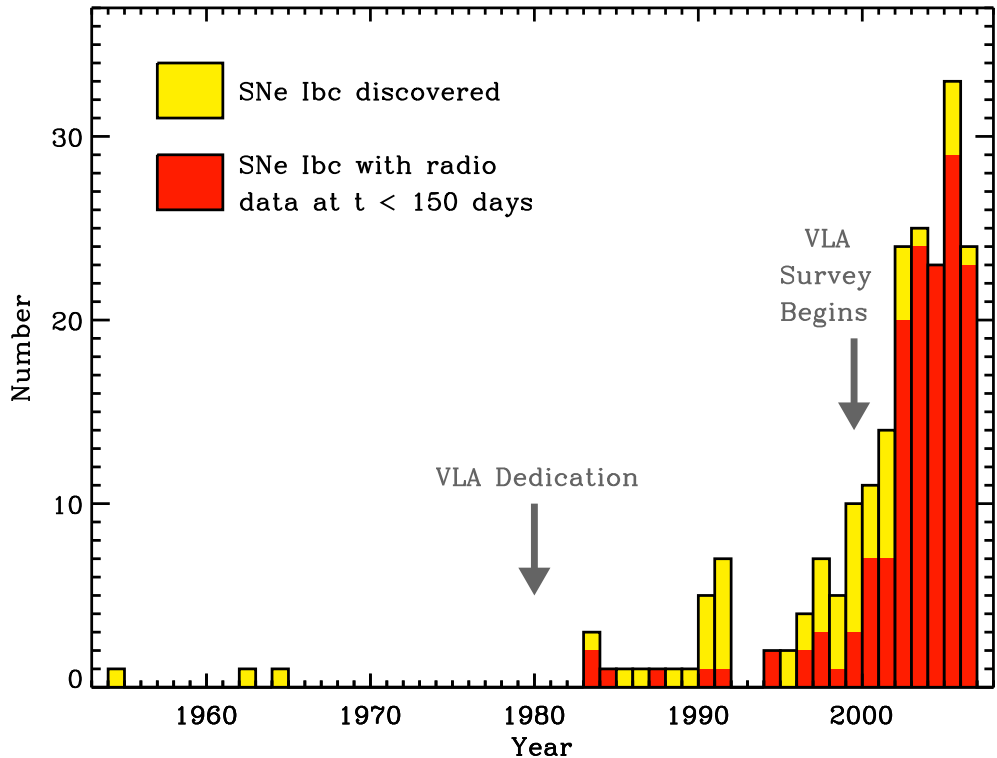}
\caption{The discovery rate of SNe Ibc each year (yellow) is compared 
with the fraction observed in the radio on timescales less than 150 
days (red).  Early radio observations are crucial for constraining
underluminous and/or off-axis GRBs since they trace the fastest
ejecta in the explosion.  Since the launch of our VLA survey, nearly
every SN Ibc has been observed on this timescale.}
\end{figure}

\section{A Large Radio Survey of SNe Ibc: The Sample}

Beginning in 2002, I have obtained radio observations for every newly
discovered SN Ibc within a maximum distance of $\sim 200$ Mpc and
accessible to the VLA.  All targets are drawn from astronomical
circulars which typically report $\sim 20$ new SNe Ibc each year, of
which $\sim 90\%$ are accessible to the VLA.  Almost all of the
reported SNe Ibc are found through ``targeted'' SN search campaigns
which monitor only the most luminous local galaxies (e.g., RC3
catalog).

Upon spectral classification, I immediately trigger a VLA
Target-of-Opportunity observation resulting in first epoch
radio observations within a few days of discovery.  Since most
SNe Ibc are discovered near maximum light, the first epoch VLA
observations correspond to days to weeks after the explosion.
Follow-up observations for each SN are scheduled logarithmically in
time since the explosion.

I supplement this sample with radio data of SNe Ibc taken before
2002.  The majority of these data were extracted from the VLA archive
and were primarily taken for radio studies of nearby galaxies, while some were
taken specifically for SN follow-up.  As a result, the archival data
generally probe significantly later timescales than my VLA survey.
This is highlighted in Figure~1 where the discovery rate of SNe Ibc is
compared with the fraction observed with the VLA;
early observations ($t\lsim 150$ days) were uncommon in the years
preceding my VLA survey.

\begin{figure}
 \includegraphics[height=.5\textheight]{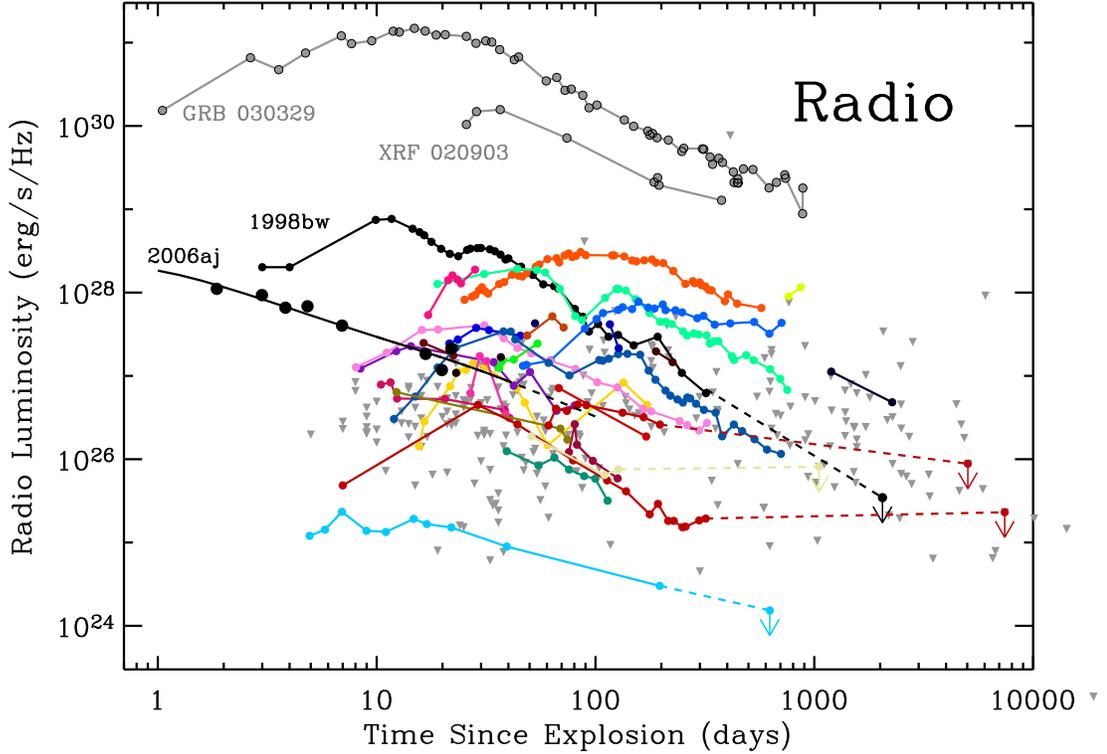}
\caption{To date, 30 local ($d\lsim 200
$ Mpc) SNe Ibc have been detected at radio wavelengths, the majority
of which were found through my dedicated VLA survey.  Detections are
shown as colored light-curves and $3\sigma$ upper limits as inverted
grey triangles.  GRB-SN\,1998bw and XRF-SN\,2006aj, also within the
maximum distance of this sample, are distinguished by their bright
early peaking radio emission (black).  Both of these events were
sub-energetic in comparison with typical GRBs (e.g., GRB\,030329;
grey) and XRFs (e.g., XRF\,020903; grey).}
\end{figure}

The combined dataset includes {\it all} VLA observations of local SNe
Ibc to date. 

\section{The Radio Properties of SNe Ibc}

This large sample of radio observations are presented in detail
in \citet{sod07}.  The large majority of these observations were 
conducted at 4.86 and/or 8.46 GHz.
As shown in Figure~2, the fraction of SNe Ibc
with detectable radio emission is small, $\sim 15\%$, corresponding
to a total of 30 SNe Ibc detected to date.  
  
\begin{figure}
 \includegraphics[height=.6\textheight]{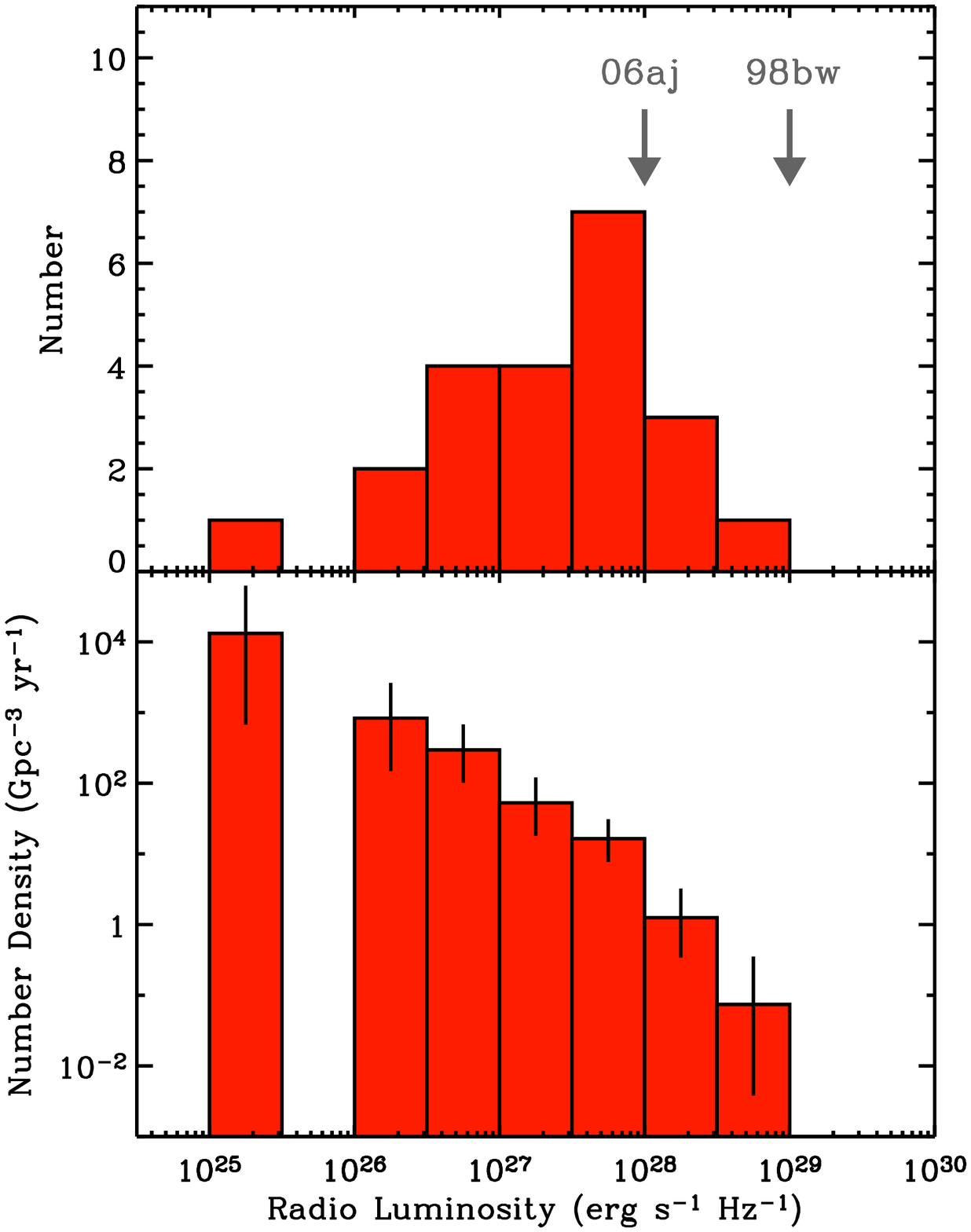}
\caption{Top: The
  peak radio luminosity distribution for local SNe Ibc peaks near
  $L_{\nu}\approx 10^{28}~\rm erg~s^{-1}~Hz^{-1}$. The peak radio
  luminosities of GRB-SN\,1998bw and XRF-SN\,2006aj are shown for
  comparison (arrows) since they lie within a comparable distance as
the optically-selected sample.
  Bottom: The radio luminosity function of SNe Ibc. Low luminosity SNe
  clearly dominate the intrinsic sample, though they are rarely
  observed since they fall below the current VLA detection limits.  We
  emphasize that this analysis includes several biases, including
  effective monitoring time pre-2002 and efficiency of optical
  spectroscopic classification.}
\end{figure}


Of those with positive detections, the majority ($\sim 70\%$) are
discovered before radio maximum with well-constrained spectral peaks.
As shown in Figures~2 and 3, the peak radio luminosities span four
orders of magnitude, $L_{\nu,\rm radio}\approx 10^{26}-10^{30}~\rm
erg~s^{-1}~Hz^{-1}$ on timescales of 1 to 1000 days.  Figure~2 shows
that GRB/XRF-SNe are distinguished from optically-selected SNe Ibc by
their strong, early-peaking radio emission.  As discussed in the next
section, these properties can be used as a proxy for the presence of
relativistic ejecta.  

Normalizing by the distance to which each SN could be detected
and the effective monitoring time of the survey, we produce the radio
luminosity function shown in Figure~3.  Clearly, sub-luminous radio SNe Ibc
are the most common, though least often detected.

As discussed in detail in \citet{sod07}, I find no evidence for any correlation
between basic optical properties (peak luminosity, photospheric velocities)
and radio luminosity.  This holds, in particular, for the fraction of 
broad-lined SNe Ibc for which the radio detection rate is no higher than that
observed for ordinary SNe Ibc.  I therefore emphasize that neither
optical luminosity nor BL spectral features are reliable proxies for
strong radio emission. 

\begin{figure}
 \includegraphics[height=.5\textheight]{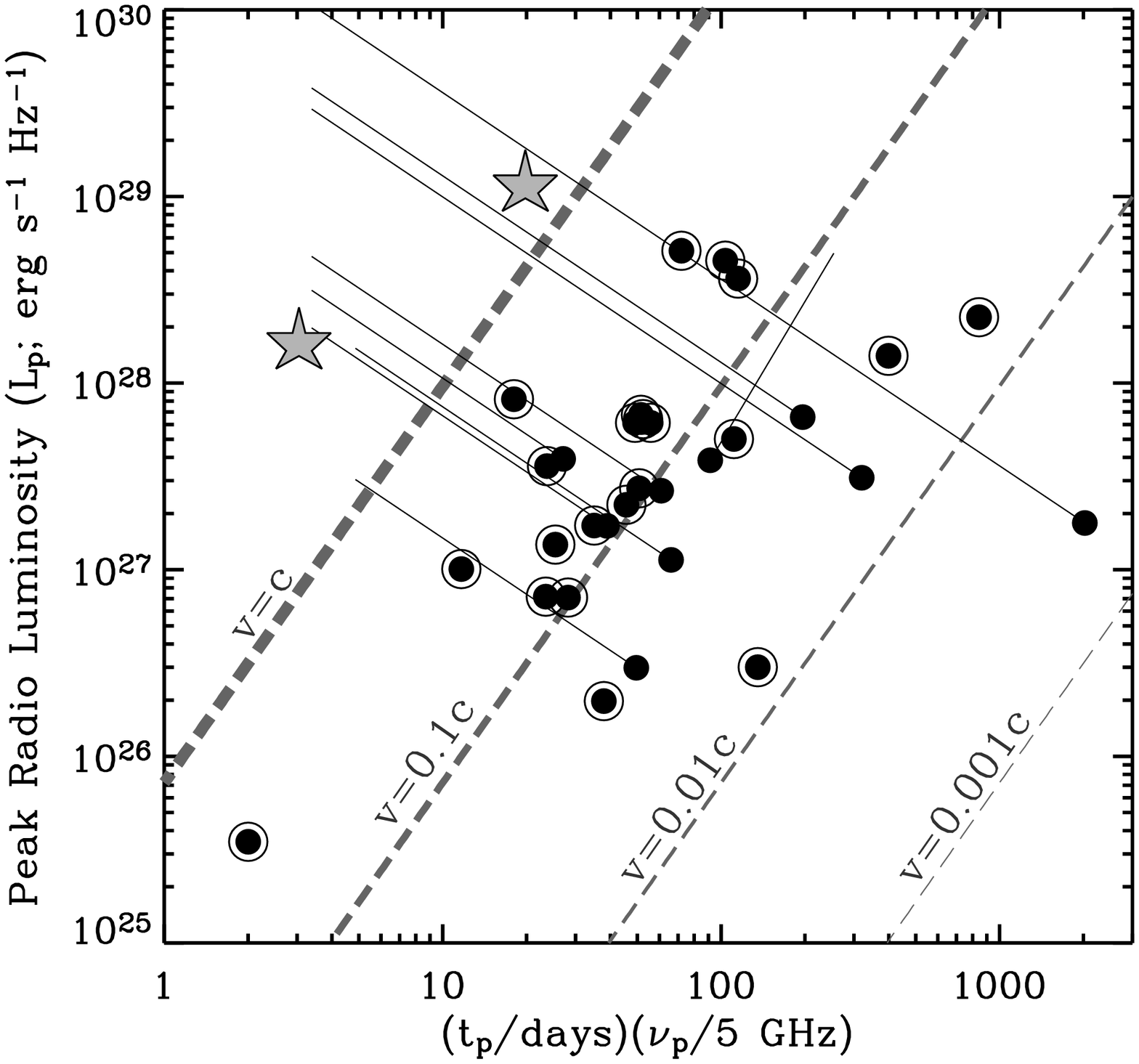}
\caption{The peak radio
  luminosities of all the SNe Ibc detected to date are
  plotted against the timescale of peak.  SNe with well-constrained
  peaks are shown as black encircled dots while those where the peak
  is not well constrained are shown as black dots with lines
  indicating the extrapolation of the radio light-curves between the
  range of typically observed peak times.  GRB-SN\,1998bw and XRF-SN\,2006aj
were also discovered within this volume and are shown as grey stars. 
Dashed lines indicate how the velocity of the fastest ejecta scales with the
observed spectral parameters.}
\end{figure}

\section{The Velocity and Energy of the Fastest Ejecta}

As discussed by \citet{c98}, the radio emission from SNe Ibc is
produced as the fastest ejecta shock-accelerate particles in the
circumstellar medium (CSM).  Turbulence amplifies the magnetic field
and the accelerated (relativistic) electrons produce synchrotron
radiation which peaks in the radio band just after the
explosion.  The spectral peak, defined by a low-frequency turn-over,
is dominated by synchrotron self-absorption (SSA) as shown by our
detailed studies of several SNe Ibc \cite{skb+05,sck+06}.  This is different
than the case for most Type II SNe which are dominated by external free-free
absorption due to a dense CSM \cite{wpm+02}.

As shown by \citet{rea94}, for radio sources dominated by SSA the
brightness temperature, $T_B=c^2 F_{\nu}/(2\pi k \theta^2 \nu^2)$, is
constrained to $5\times 10^{10}$ K, assuming the post-shock energy
density is in equipartition between magnetic fields, $\epsilon_B$, and
relativistic electrons, $\epsilon_e$.  With this assumption, robust
constraints on the radius, $r$, and internal energy of the shocked
CSM, $E_i$, are derived \citep{c98,kfw+98,cf06}.  These quantities
scale as simple observables including the peak spectral
frequency, $\nu_p$, the flux at the spectral peak, $f_p$, and the
luminosity distance, $d$ (see \citealt{skb+05,sck+06} for a detailed
discussion). Here we adopt the simple model of \citealt{cf06},
assuming $\epsilon_e=\epsilon_B=0.1$ which provides the following
relations:

\begin{equation}
r =  4\times 10^{14} (f_p/{\rm mJy})^{9/19} (d/{\rm Mpc})^{18/19} (\nu_p/{\rm 5~
GHz
})^{-1}~{\rm cm}
\end{equation}
\begin{equation}
E_i  =  1.5\times 10^{41} (d/{\rm Mpc})^8 (f_p/{\rm mJy})^4 (\nu_p/{\rm 5~GHz})^
{-7} (r/10^{15}\rm~cm)^{-6}~\rm erg.
\end{equation}

\noindent
We note that the radius of the emitting material is only weakly
dependent on the energy partition fractions, while departures from
equipartition produce a strong increase the energy.  Taken together
with the observed peak time, $t_p$, the velocity of the shock is
easily estimated from Equation~1 and the resulting values are shown
for these SNe Ibc in Figure~4.  SNe with early, bright radio emission
have the fastest ejecta.  As shown in the Figure, of the thirty radio
bright SNe Ibc in this sample, the inferred velocities range from 0.01
to 0.5c, and none show the mildly-relativistic ejecta inferred for
GRB-SN\,1998bw \cite{kfw+98} or XRF-SN\,2006aj \cite{skn+06}.

\begin{figure}
 \includegraphics[height=.5\textheight]{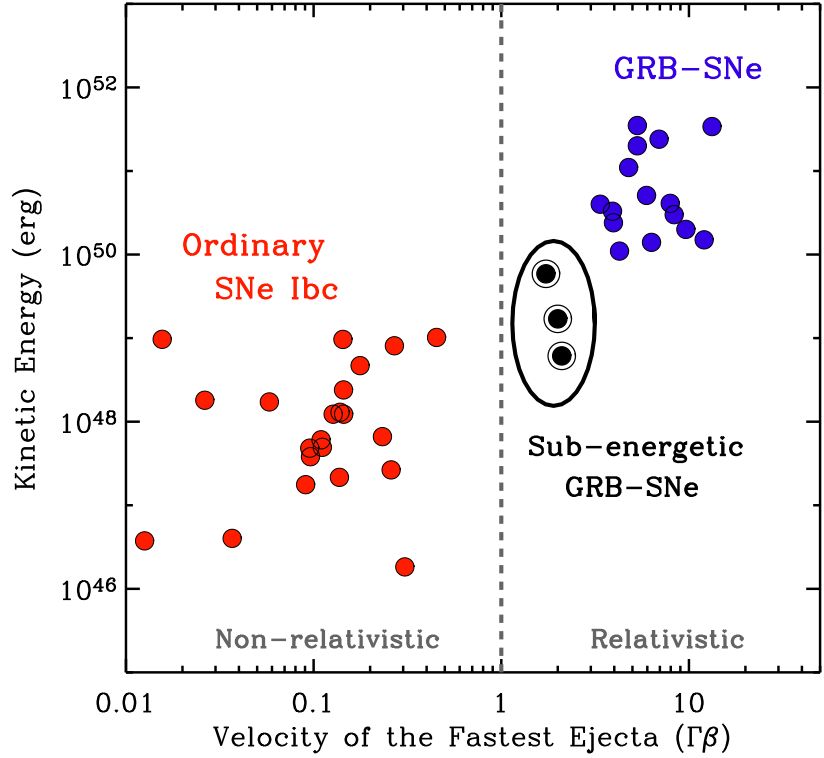}
\caption{The kinetic
  energy and velocity of the fastest ejecta for radio bright SNe Ibc
  are compared with those of GRBs, XRFs, and sub-energetic bursts.
  Ordinary SNe Ibc (red) are distinguished in that they couple
  $10^{48}$ erg to non-relativistic ejecta at $v\approx 0.15c$ while
  GRBs and XRFs (blue) couple at $10^{51}$ erg to relativistic
  material ($\Gamma\approx 10$).  Sub-energetic bursts (black) are
  intermediate between the two classes, coupling at least $10^{48}$
  erg to mildly-relativistic ($\Gamma\approx 3$).}
\end{figure}

Next is a discussion of the explosion energetics.  As derived by \citet{c83},
$E_i$ is just $\sim 20\%$ of the total energy (kinetic and internal)
of the shocked CSM, which in turn is equal to the kinetic energy of
the fastest ejecta.  Including this conversion factor, we find that
the kinetic energy of the fastest ejecta, $E_K\sim 5E_i$, spans
$10^{46}-10^{49}$ erg for SNe Ibc, a factor of $10^2$ to $10^5$ times
less than that traced by the optical emission.

Finally, we compare these ejecta parameters (velocity, kinetic energy)
with those of GRB and XRFs, including the class of sub-energetic
bursts (Figure~5).  We find evidence of a clear dichotomy
between ordinary SNe Ibc and typical GRBs/XRFs: ordinary SNe couple
roughly $10^{48}$ erg to their fastest (but non-relativistic)
ejecta at $v\sim 0.15c$.  Meanwhile, typical GRBs/XRFs couple $10^{51}$
erg to relativistic jets with $\Gamma\sim 10$.  Sub-energetic explosions
bridge these two classes, coupling at least $10^{48}$ erg to
mildly-relativistic outflows with $\Gamma\sim 3$.

A related question is whether the observed radio emission for some
SNe Ibc may be suppressed due to viewing angle effects, specifically
important in the case of an associated GRB jet directed away from our
line-of-sight \cite{pac01}.  In this scenario a rapid increase in
radio emission is expected as the jet decelerates and spreads
sideways, eventually crossing our line-of-sight on a timescale of
$\sim 1$ yr after the explosion.  As shown in Figure~2 and discussed in 
\citet{snb+06}, radio observations of $\sim 70$ SNe Ibc taken between months to
decades after the explosion reveal no evidence for associated GRB
jets.  Statistically we constrain the fraction of SNe Ibc harboring
typical GRB jets to be less than $10\%$.

\section{Conclusions}

In conclusion, based on our large survey of optically selected local
SNe Ibc, I find that (i) roughly 15\% show detectable radio emission,
(ii) the peak radio luminosities of SNe Ibc span four orders of
magnitude with peak times between 1 and 1000 days, (iii) the kinetic
energy and fastest velocity of these explosions are significantly
different from those of typical GRBs and XRFs, (iv) compared with the
sample of sub-energetic GRBs, $\lsim 3\%$ of SNe Ibc show ejecta
properties indicative of an engine-driven explosion, and (v) 
the fraction of SNe Ibc harboring classical GRB jets 
viewed off-axis is constrained to less than 10\%. 

Despite this progress, it remains an open question why just $\sim 1$\%
of SNe Ibc give rise to GRBs or XRFs.  A deeper understanding of the
progenitors of ordinary SNe Ibc will shed light on this issue.  With
the recent advent of wide-field optical surveys, SNe Ibc from ``blind''
search campaigns will soon dominate new discoveries.  Any
investigation of the relation between the large-scale environment
(host galaxy characteristics) and ejecta properties relies on SNe
discovered through these unbiased surveys. In particular, host galaxy
metallicity is argued to be a critical parameter as a proxy for the
progenitor metallicity.  Numerical models suggest that only low
metallicity progenitors are able to produce the accretion-disk powered
outflows inferred for GRBs \citep{whw02}.  Therefore, these models
predict that engine-driven explosions are unlikely to be hosted in the
luminous, high-metallicity galaxies monitored by targeted SN
searches \citep{sgb+06}.  Looking forward, a radio survey focused
exclusively on the SNe Ibc discovered through blind surveys will
directly address and answer these crucial questions.





\bibliographystyle{aipproc}   


\begin{thebibliography}{20}
\expandafter\ifx\csname natexlab\endcsname\relax\def\natexlab#1{#1}\fi
\providecommand{\enquote}[1]{``#1''}
\expandafter\ifx\csname url\endcsname\relax
  \def\url#1{\texttt{#1}}\fi
\expandafter\ifx\csname urlprefix\endcsname\relax\def\urlprefix{URL }\fi
\providecommand{\eprint}[2][]{\url{#2}}

\bibitem[{Woosley} and {Bloom}(2006)]{wb06}
S.~E. {Woosley}, and J.~S. {Bloom}, \emph{ARA\&A} \textbf{44}, 507--556 (2006).

\bibitem[{Soderberg} et~al.(2006{\natexlab{a}})]{skp+06}
A.~M. {Soderberg}, {\it et al.}, \emph{ApJ} \textbf{636}, 391--399 (2006{\natexlab{a}}).

\bibitem[{Pian} et~al.(2006)]{pmm+06}
E.~{Pian}, {\it et al.}, \emph{Nature} \textbf{442}, 1011--1013 (2006).

\bibitem[{Soderberg} et~al.(2004{\natexlab{a}})]{skb+04b}
A.~M. {Soderberg}, {\it et al.}, \emph{{Nature}} \textbf{430}, 648--650
  (2004{\natexlab{a}}).

\bibitem[{Soderberg} et~al.(2006{\natexlab{b}})]{skn+06}
A.~M. {Soderberg}, {\it et al.}, \emph{Nature} \textbf{442}, 1014--1017 (2006{\natexlab{b}}),

\bibitem[{Berger} et~al.(2003)]{bkf+03}
E.~{Berger}, S.~R. {Kulkarni}, D.~A. {Frail}, and A.~M. {Soderberg}, \emph{ApJ}
  \textbf{599}, 408--418 (2003).

\bibitem[{Soderberg} et~al.(2004{\natexlab{b}})]{sfw04}
A.~M. {Soderberg}, D.~A. {Frail}, and M.~H. {Wieringa}, \emph{ApJL}
  \textbf{607}, L13--L16 (2004{\natexlab{b}}).

\bibitem[{Soderberg} et~al.(2006{\natexlab{c}})]{snb+06}
A.~M. {Soderberg}, E.~{Nakar}, E.~{Berger}, and S.~R. {Kulkarni}, \emph{ApJ}
  \textbf{638}, 930--937 (2006{\natexlab{c}}).

\bibitem[{Soderberg}(2007)]{sod07}
A.~M. {Soderberg}, \emph{ApJ}  (2007), in preparation.

\bibitem[{Chevalier}(1998)]{c98}
R.~A. {Chevalier}, \emph{ApJ} \textbf{499}, 810--+ (1998).

\bibitem[{Soderberg} et~al.(2005)]{skb+05}
A.~M. {Soderberg}, {\it et al.}, \emph{ApJ} \textbf{621}, 908--920
  (2005).

\bibitem[{Soderberg} et~al.(2006{\natexlab{d}})]{sck+06}
A.~M. {Soderberg}, R.~A. {Chevalier}, S.~R. {Kulkarni}, and D.~A. {Frail},
  \emph{ApJ} \textbf{651}, 1005--1018 (2006{\natexlab{d}}).

\bibitem[{Weiler} et~al.(2002)]{wpm+02}
K.~W. {Weiler}, N.~{Panagia}, M.~J. {Montes}, and R.~A. {Sramek}, \emph{ARA\&A}
  \textbf{40}, 387--438 (2002).

\bibitem[{Readhead}(1994)]{rea94}
A.~C.~S. {Readhead}, \emph{ApJ} \textbf{426}, 51--59 (1994).

\bibitem[{Kulkarni} et~al.(1998)]{kfw+98}
S.~R. {Kulkarni}, {\it et al.}, \emph{Nature} \textbf{395}, 663--669 (1998).

\bibitem[{Chevalier} and {Fransson}(2006)]{cf06}
R.~A. {Chevalier}, and C.~{Fransson}, \emph{ApJ} \textbf{651}, 381--391 (2006).

\bibitem[{Chevalier}(1983)]{c83}
R.~A. {Chevalier}, \emph{ApJ} \textbf{272}, 765--772 (1983).

\bibitem[{Paczynski}(2001)]{pac01}
B.~{Paczynski}, \emph{Acta Astronomica} \textbf{51}, 1--4 (2001).

\bibitem[{Woosley} et~al.(2002)]{whw02}
S.~E. {Woosley}, A.~{Heger}, and T.~A. {Weaver}, \emph{Reviews of Modern
  Physics} \textbf{74}, 1015--1071 (2002).

\bibitem[{Stanek} et~al.(2006)]{sgb+06}
K.~Z. {Stanek}, {\it et al.}, \emph{Acta Astronomica} \textbf{56}, 333--345 (2006).

\end{thebibliography}




\end{document}